\documentclass[prd, twocolumn, superscriptaddress,floatfix, nofootinbib, preprintnumbers]{revtex4-2}
\usepackage[utf8]{inputenc}
\usepackage[colorlinks=true,citecolor=blue]{hyperref}
\usepackage{amssymb}
\usepackage{amsmath}
\usepackage{graphicx}
\usepackage{footmisc}
\usepackage{url}
\usepackage{multirow}
\usepackage{makecell}
\usepackage{hhline}
\usepackage{comment}
\usepackage{array}
\usepackage{float}
\usepackage{ragged2e}
\usepackage{threeparttable}
\usepackage{enumitem}

\usepackage{xcolor}

\def\beq{\begin{equation}}
\def\eeq{\end{equation}}
\newcommand{\bea}{\begin{eqnarray}\begin{aligned}}
\newcommand{\eea}{\end{aligned}\end{eqnarray}}

\newcommand{\mycomment}[1]{}

\newcommand{\vdata}{v^{\rm data}_{\theta}}
\newcommand{\vCR}{v^{\rm CR}_{\theta}}
\newcommand{\vtildata}{\tilde{v}^{\rm data}_{\theta}}

\newcommand{\vint}{v^{\rm int}_{\theta}}

\begin{document}

\title{SIGMA: Single Interpolated Generative Model for Anomalies}

\author{Ranit Das}
\email{ranit@physics.rutgers.edu}
\affiliation{NHETC, Dept.\ of Physics and Astronomy, Rutgers University, Piscataway, NJ 08854, USA}

\author{David Shih}
\email{shih@physics.rutgers.edu}
\affiliation{NHETC, Dept.\ of Physics and Astronomy, Rutgers University, Piscataway, NJ 08854, USA}

\begin{abstract}

A key step in any resonant anomaly detection search is accurate modeling of the background distribution in each signal region.
Data-driven methods like CATHODE accomplish this by training separate generative models on the complement of each signal region, and interpolating them into their corresponding signal regions. Having to re-train the generative model on essentially the entire dataset for each signal region is a major computational cost in a typical sliding window search with many signal regions. Here, we present SIGMA, a new, fully data-driven, computationally-efficient method for estimating background distributions. The idea is to train a single generative model on all of the data and interpolate its parameters in sideband regions in order to obtain a model for the background in the signal region. The SIGMA method significantly reduces the computational cost compared to previous approaches, while retaining a similar high quality of background modeling and sensitivity to anomalous signals.
 
\end{abstract}

\maketitle

\section{Introduction}

There has been a rapid growth in new methods for resonant anomaly detection in recent years~\cite{Collins:2018epr,Collins:2019jip,anode,Andreassen:2020nkr,Stein:2020rou,Amram:2020ykb,cathode,Collins:2021nxn,1815227,Kasieczka:2021tew,Hallin:2022eoq,Chen:2022suv,Kamenik:2022qxs,Sengupta:2023xqy,Raine:2022hht,Golling:2023yjq, feta,bickendorf2023combining, cathodebdt,anode_bdt,full_AD,ranode,Beauchesne:2023vie,drapes, lhco_paper, curtrainsf4f, Mastandrea:2022vas, Cheng:2024yig, Benkendorfer_2021,radot,sophon}. These methods are starting to move from proof-of-concept to actual analyses applied to ATLAS \cite{ATLAS:2020iwa} and CMS data \cite{CMS-PAS-EXO-22-026} in the model-agnostic search for new physics. They have also been successfully applied to astronomical data from the Gaia Space Telescope in the search for stellar streams \cite{Shih_2021, shih2023machinae20fullskymodelagnostic,sky_cwola, skycurtains}. 

In resonant anomaly detection, one uses additional features $x$ to enhance the sensitivity of a ``bump hunt" in a primary feature $m$ where the signal is assumed to be localized. The overall goal is to accurately estimate the Neyman-Pearson classifier
\begin{equation}
\label{eq:Roptimal}
R_{\rm optimal}(x) = {p_{\rm data}(x)\over p_{\rm background}(x)}
\end{equation}
in a signal region (SR), which is defined as a window in $m$. By cutting on this ``optimal anomaly score", one can preferentially select signal over background events in a model-agnostic way, i.e.\ without having to specify the signal model in advance. 

In many approaches, a key step is to learn a fully data-driven estimate for $p_{\rm background}(x)$ by interpolating from from neighboring control regions (CR) in $m$. In methods like ANODE~\cite{anode} and CATHODE~\cite{cathode}, one trains a generative model and/or density estimator (e.g.\ a conditional normalizing flow) on the complement of the signal region, which for a narrow signal region is almost the entire dataset. For each signal region this must be performed separately, and this leads to extremely expensive trainings for a full sliding-window search.

There have been efforts to mitigate this computational cost. In CURTAINSF4F~\cite{curtrainsf4f}, one can train a single flow on the entire dataset, not masking out the SR, and another conditional flow on short control regions to learn the transformation between them. This can then be used to interpolate into the signal region and saves significantly on computational cost since the all-data flow need only be learned once, and the short control region flow is usually learned on significantly less data.

Recently in RAD-OT~\cite{radot}, the idea was to dispense with the flow altogether and just use OT to transport sidebands to sidebands, thereby interpolating into the SR. This was shown to be computationally very inexpensive but at the cost of sacrificing some quality in the interpolation.

In this paper, we present a Single Interpolated Generative Model for Anomalies (SIGMA), a new approach for computationally efficient resonant anomaly detection, which is much simpler than CURTAINSF4F and maintains the high quality of the more expensive methods like CATHODE. Our idea is to train a single flow on all the data, as in CURTAINSF4F, and simply interpolate the parameters of the flow in the resonant feature. We find the simplest implementation of this is using flow-matching, where the vector field itself can be easily interpolated into the signal region. We will demonstrate that SIGMA produces high quality background samples while saving significantly on training cost -- even more than CURTAINSF4F since even the short sideband flows can be dispensed with. 

This paper is organized as follows: in Section~\ref{sec:dataset} we describe the dataset used to demonstrate our method. The architecture of the generative model used for our background template is described in ~\ref{sec:network}. In \ref{sec:interpolation_CR} we describe the slower interpolation method from \cite{anode,cathode} using flow-matching. The new interpolation method SIGMA is described in Section~\ref{sec:interpolation_new}. The model hyperparameters and the training are described in Section~\ref{sec:model_hyperparams}. In Section~\ref{sec:results}, we show the results for anomaly detection, and the background template obtained from different methods. Finally in Section~\ref{sec:conclusions} we present our conclusions and further discussion. In Appendix~\ref{appendix:freq_embedding}, we discuss the importance of using frequency embedding for SIGMA. In Appendix~\ref{appendix:classifier_test} we analyze the background templates using the classifier test, developed in \cite{das2023understandlimitationsgenerativenetworks}. In Appendix~\ref{appendix:sample_quality_SR}, we discuss the background template quality for SIGMA for different Signal Regions.

\section{Setup}
\label{sec:setup}

\subsection{Dataset}
\label{sec:dataset}
We use the LHCO R\&D Dataset \cite{lhco, Kasieczka:2021xcg} for this study with 
dataset and train-val-test splits similar to \cite{cathode,anode}. The Standard Model(SM) background consists of QCD dijet events, and $W' \rightarrow X(\rightarrow qq) Y(\rightarrow qq)$ events with $m_{W'}=3.5\,\text{TeV}$, $m_{X}=500\,\text{GeV}$ and $m_{Y}=100\,\text{GeV}$ form the signal. These are simulated using \texttt{Pythia 8} \cite{pythia_1,pythia_2} and \texttt{Delphes 3.4.1} \cite{delphes_1, delphes_2, delphes_3}. The reconstructed particles are clustered into jets using the anti-$k_T$ algorithm \cite{antikt_1, antikt_2} with $R=1$ using \texttt{Fastjet} \cite{fastjet}. Events that satisfy the $p_T > 1.2\,\text{TeV}$ jet trigger are kept. 

\mycomment{We use 2 sets of training features: 
\begin{enumerate}
    \item Baseline features
    \begin{equation}
 m=m_{JJ},\quad    x =[m_{J_1}, \Delta m_J, {\tau_{21}}^{J_1},{\tau_{21}}^{J_2}],
\end{equation} 
    \item Baseline features with $\Delta R$
    \begin{equation}
 m=m_{JJ},\quad    x =[m_{J_1}, \Delta m_J, {\tau_{21}}^{J_1},{\tau_{21}}^{J_2}, \Delta R],
\end{equation} 
\end{enumerate}}

The training features we use are:
\begin{equation}
 m=m_{JJ},\quad    x =[m_{J_1}, \Delta m_J, {\tau_{21}}^{J_1},{\tau_{21}}^{J_2}, \Delta R],
\end{equation} 
where invariant masses of the subjets satisfy $m_{J_1}<m_{J_2}$, and $\Delta
m_J = m_{J_2}-m_{J_1}$. We chose dijet invariant mass $m_{JJ}$ as the resonant variable $m$, with the signal region (SR) defined as $m \in [3.3,3.7]$ and  $m \notin [3.3,3.7]$ form the control regions (CR). The n-subjettiness ratios are $\tau_{ij} = \tau_i/\tau_j$ \cite{nsub_1,nsub_2}. $\Delta R$ is the angular separation of the leading two jets in $\eta - \phi$ space. Including it does not enhance the performance for anomaly detection, compared to just baseline features. But it is a feature that is strongly correlated with the resonant feature, $m_{JJ}$, and hence including it provides a stringent test for any interpolation method~\cite{curtains}.   

We use all $1$-million SM background events from the original R\&D dataset and inject different amounts of signal (3000 or lower) from the first 70k signal events of the R\&D dataset. This is meant to represent actual un-labeled data from the experiment.  For reference, 
the signal-injection of $N_{\rm sig}=1000$ corresponds to $\approx 2.2\sigma$ significance for the inclusive bump hunt in $m_{JJ}$.

\begin{figure*}
    \includegraphics[width=0.78\textwidth]{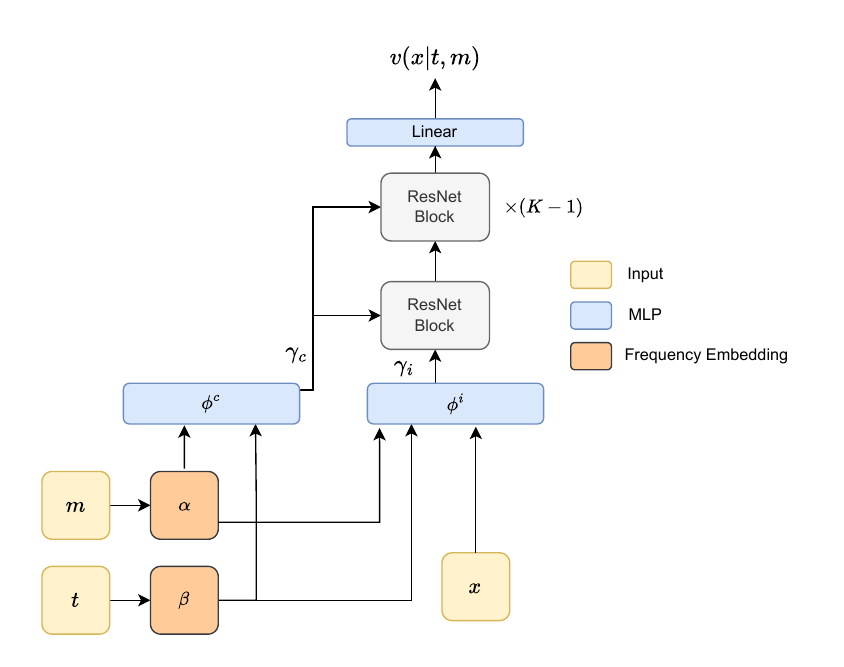}
    \caption{ResNet architecture}
    \label{fig:resnet}
\end{figure*}

We split this dataset into training and validation sets, with 85-15 split for training all generative models. For IAD, following \cite{cathode}, we classify between data in SR and an additional 272k QCD dijet events in the SR \cite{extra_qcd} (which represents true background). For CATHODE, we classify between data in SR, and 360k events sampled from the 10 best instances of the trained generative model. We use the remaining 30k signal events from the original R\&D dataset, along with the additional 340k QCD dijet events in the SR from \cite{extra_qcd} for evaluating SIC curves. 

For each $N_{\rm sig}$, we produce ten different datasets by injecting different randomly selected signal events from the R\&D dataset. This is used to obtain the error bars on the performance curves in the figures.
\subsection{Flow-matching and Network Architecture} 
\label{sec:setup_network}
We use Conditional Flow Matching~\cite{lipman2023flowmatchinggenerativemodeling, birk2023flowmatchingkinematicsgenerating, buhmann2023epiclyfastparticlecloud} to model the background template as a proof of concept, but our idea could be extended to be used with Diffusion models~\cite{ho2020denoisingdiffusionprobabilisticmodels} as well. Besides providing a better quality background template than ordinary normalizing flows, we found that flow matching also trains faster -- in approximately 30 mins, as compared to a normalizing flow which trains in $\sim$ 2.8 hours. In this section, we give a brief description of conditional flow-matching and then describe the neural network architecture used to model the flow.

\subsubsection{Conditional Flow Matching}

Conditional Flow Matching (CFM) \cite{lipman2023flowmatchinggenerativemodeling,ot_cfm_2, fm_4, fm_5, fm_6} is an efficient way to train Continuous Normalizing Flows (CNFs) \cite{cnf}. In CNFs, one aims to learn the vector field $u_t(x_t): [0,1] \times \mathbb{R}^d \to \mathbb{R}^d$ which implements a continuous transformation of data $x_t$ via:
\begin{equation}
\label{eq:vector_field}
    \frac{d x_t}{dt} = u_t(x_t),
\end{equation}
At $t=0$, $x_0$ follows the data distribution $p_\text{data}(x_0)$, and at $t=1$, $x_1$ follows a known distribution $p_{\text{base}}(x_1)$. Here we take $p_{\text{base}}(x)$ to be the unit normal distribution $\mathcal{N}(x|0,1)^d$. In between, $x_t$ is sampled from a density $p_t(x_t)$ which is generated by the vector field; the trajectory of these densities is referred to as the probability density path.  

Training CNFs with maximum likelihood is extremely computationally expensive, because each evaluation of the likelihood requires solving an ODE. The breakthrough insight of Conditional Flow Matching is that one can instead regress the desired vector field $u_t(x_t)$ with an MSE-type loss, by matching it to a {\it conditional} vector field  $u_t(x_t|x_0)$. This generates a conditional probability path $p_t(x|x_0)$ where at $t=0$, we have $p_0(x|x_0)=\mathcal{N}(x|x_0,\sigma^2 I
)$ where $\sigma^2$ is very small, and at $t=1$, we have $p_1(x|x_0) = \mathcal{N}(x|0,1)^d$.
Marginalizing this path over the data distribution $p_{\text{data}}(x_0)$ gives us the unconditional probability $p_t(x)$: 
\begin{equation}
p_t(x)=\int d x_0\, p_t(x|x_0) p_{\text{data}}(x_0).
\end{equation}
 In CFM, this conditional vector field is regressed with a neural network $v_{\theta}(x_t|t)$ by minimizing the CFM loss
\begin{equation}
\mathcal{L}(\theta) = \left\|v_{\theta}(x_t|t)-u_t\left(x_t|x_0\right)\right\|^2,
\end{equation}
In \cite{lipman2023flowmatchinggenerativemodeling}, it was shown that the CFM objective, when averaged over $t \sim \mathcal{U}[0,1]$, $x_0 \sim p_{\text{data}}(x_0)$ and $x_t \sim  p_t(x|x_0)$, is minimized by $v_{\theta}(x|t) = u_t(x)$. In addition to $t$, our models also have an additional conditional feature: $m$, the resonant feature, which allows us to model the vector field $u_t(x|m)$ corresponding to $p(x|m)$. After integrating equation \eqref{eq:vector_field} to obtain $x_t$, $p(x_0|m)$ can be obtained using:
\begin{equation}
\label{eq:flow_density}
    \log p(x_0|m) = \log(x_1)-\int_1^{0} dt \ \nabla_{x_t} . u_t(x_t|m).
\end{equation}

\subsubsection{Network Architecture}
\label{sec:network}
 Similar to \cite{drapes}, we use a \texttt{ResNet}-style \cite{resnet} architecture to model the vector field $u_t(x|m)$, since it is more expressive than an MLP. Rather than pass $m$ and $t$ directly to the \texttt{ResNet}, it is beneficial to first map them to an embedding space. For $t$, we use a frequency embedding $\beta: [0,1] \rightarrow \mathbb{R}^{2(L'+1)}$ , similar to those used in diffusion models \cite{ho2020denoisingdiffusionprobabilisticmodels}: 
    \begin{equation}
    \begin{split}
    \beta(t) = &\left(\sin(\pi t), \cos(\pi t), \dots,\right. \\
    &\left.\sin((L'+1) \pi t), \cos((L'+1) \pi t)\right).
    \end{split}
    \end{equation}
The resonant feature $m$ is first rescaled to $[-1,1]$, and then mapped to an embedding space $\alpha: [-1,1] \rightarrow \mathbb{R}^{2L}$. Depending on the interpolation method, we find different embeddings are preferable, see Sec.~\ref{sec:setup_interpolation} for more details.
The $t$ and $m$ embeddings are concatenated and passed to an MLP to produce the full context vector:
    \begin{equation}\label{eq:cond_embed}
        \gamma_c(m,t) = \phi_c(\alpha(m),\beta(t)),
    \end{equation}
    where $\phi_c:\mathbb{R}^{2 (L+L')} \to \mathbb{R}^n$ is an \texttt{MLP}. 
    
    The features $x \in \mathbb{R}^d$ are also concatenated with the frequency embedding for time and conditional features to form the input vector $\gamma_i(x,m,t)$:
    \begin{equation}\label{eq:input_embed}
        \gamma_i(x,m,t) = \phi_i(x,\alpha(m),\beta(t)),
    \end{equation}
    where $\phi_i: \mathbb{R}^{2 (L+L')+d} \to \mathbb{R}^n $ is another \texttt{MLP}.
    
    Finally, the context and input vectors are passed through $K$ \texttt{ResNet} blocks as shown in Figure~\ref{fig:resnet}. 
    The $k^\text{th}$-\texttt{ResNet} Block $h^k:\mathbb{R}^n \times \mathbb{R}^n \to \mathbb{R}^n $ takes as input $h^{k-1}$ and the context vector $\gamma_c$, and outputs
    \begin{equation}
    \label{eq:resnet_conditioning}
        h^k = h^{k-1} + \phi^k(h^{k-1}) * \gamma_c,
    \end{equation}
    where $\phi^k:\mathbb{R}^n \to \mathbb{R}^n $ is the MLP corresponding to the \texttt{ResNet} block. For $k=0$, $h^0=\gamma_i$.
    
    We use the \texttt{ResNet} model defined in the \texttt{nflows} package \cite{nflows}. The output of the final \texttt{ResNet} block $h^K$ is passed through a linear layer (to change dimensions from the $n$-dimensional input vector space to the $d$-dimensional feature space) in order to obtain the final vector field $v_{\theta}(x|t,m)$.

    \mycomment{
    The first \texttt{ResNet} Block is defined as $h^1: \mathbb{R}^n \times \mathbb{R}^n \rightarrow \mathbb{R}^n$:
    \begin{equation}\label{eq:\texttt{ResNet}1}
        h^1(\gamma_i, \gamma_c) = \phi^1(\gamma_i) * \gamma_c + \gamma_i,
    \end{equation} and the $k^{th}$ \texttt{ResNet} Block $h^k: \mathbb{R}^n \times \mathbb{R}^n \rightarrow \mathbb{R}^n$ is defined recursively as :
    \begin{equation}\label{eq:\texttt{ResNet}2}
        h^k = (\phi^k \circ h^{k-1}) * \gamma_c + \gamma_c, \quad k=2, \dots, K,
    \end{equation}
    where $\phi^k:\mathbb{R}^n \to \mathbb{R}^n$ is an \texttt{MLP} and $*$ denotes component-wise multiplication.}

\subsection{Interpolation}
\label{sec:setup_interpolation}
 
In this section we discuss the two different types of interpolation: First we discuss a flow-matching version of background interpolation used in \cite{cathode, anode, lacathode, ranode}. Then we discuss our main result, the SIGMA method, where we first train a model, $v^{\text{data}}_{\theta}$ to learn all data, and then obtain $v^{\text{int}}_{\theta}$ for each SR through a fast interpolation procedure.

\mycomment{
\begin{figure*}
    \centering
    \includegraphics[width=0.48\textwidth]{sic_signal_scan.pdf}

    \caption{SIC at $\epsilon_B=0.001$ for different signal injections for (Left) the baseline dataset, (Right) the baseline dataset with $\Delta R$. We see that the SIC values at this working point are better for the data model with interpolation, as compared to without interpolation. This difference is much more pronounced for the case with $\Delta R$}.
    \label{fig:signal_scan}
\end{figure*}}

\subsubsection{Interpolation using $v_{\theta}^{\text{CR}}(x|t,m)$}
\label{sec:interpolation_CR}

Similar to \cite{anode,cathode}, we use $\alpha(m) = m,$ for the embedding for the resonant feature. 
For every SR, a model is trained to learn $v_{\theta}^{\text{CR}}(x|t,m \in {\rm CR})$ for its corresponding CR data. Since one needs to train a separate $v_{\theta}^{\text{CR}}(x|t,m)$ for each CR, this method can be computationally expensive. Once trained, the interpolated vector field into SR is simply $v_{\theta}^{\text{CR}}(x|t,m \in \text{SR})$. Since the CR-only training did not ever see the signal in the SR, the interpolation into the SR can't have learned the signal. Therefore the CR-only interpolation method should give a model for the background-only distribution.

\mycomment{For this method, we use the same base \texttt{ResNet} model but without a frequency embedding for the resonant feature, as was discussed in Section~\ref{sec:network}.}

\subsubsection{SIGMA: interpolation using $v_{\theta}^{\text{data}}(x|t,m)$}
\label{sec:interpolation_new}

In SIGMA, the idea is to first learn the full data distribution (including the signal region) as well as possible,\footnote{One could
attempt to deliberately make $\vdata$ less expressive, in order to not learn the signal in the SR, and protect the interpolation against signal contamination. But this would be difficult to control and likely would result in a worse fit to the background overall.} and then to interpolate the {\it parameters} of the vector field from $m$ outside the SR into $m$ into the SR, so that the interpolated background model does not include signal. Since the interpolation step doesn't require any additional training time, the background template for different SRs can be obtained trivially from the single, common data training. 

To learn the full data distribution optimally, including the more localized, higher frequency modes corresponding to signal, we found that a frequency embedding for $m$ was beneficial \cite{nerf, spectral_bias}.
The resonant feature $m$ is first re-scaled to $(-1,1)$, and then $\alpha(m)$ is defined as:
\begin{equation}
\label{eq:mass_embed}
\begin{split}
\alpha(m) = &\left(\sin(2^{0} \pi m), \cos(2^{0} \pi m), \dots, \right.\\
&\left.\sin(2^{L-1} \pi m), \cos(2^{L-1} \pi m)\right).
\end{split}
\end{equation}

There are in principle many ways to interpolate the vector field from the CR into the SR. We will compare two in this work and see which one works better empirically. In real-world applications where there is no access to truth labels (background vs.\ signal), one could still determine the better interpolation method using simulations, or using signal injection tests on top of real data, similar to how ATLAS and CMS currently set limits on searches using weak supervision \cite{ATLAS:2020iwa,CMS-PAS-EXO-22-026}.

The two interpolation methods will explore are:
\begin{enumerate}
\item Linearly interpolating the vector field itself:
\bea
\label{eq:int_linear}
& v_\theta^{\text{int}}(x|t,m)= \\
&\quad \xi * \,v_\theta^{\text{data}}(x|t,m_1) +
(1-\xi) * v_\theta^{\text{data}}(x|t,m_2)
\eea
where $\xi = (m-m_2)/(m_1 - m_2)$, with $m \in \text{SR}$, $m_1, m_2 \in \rm CR$, and $m_2 <m < m_1$.

\item
Linearly interpolating the context and input vectors:
\bea
\label{eq:int_embed}
         \gamma_c^{\text{int}}(m,t) &= \xi * \gamma_c(m_1,t) + (1-\xi) * \gamma_c(m_2,t) \\
        \gamma_i^{\text{int}}(x,m,t) &= \xi * \gamma_i(x,m_1,t) \\
&\qquad\qquad + (1-\xi) * \gamma_i(x,m_2,t),
\eea
And then the interpolated vector field $\vint$ is just given by $v_\theta^{\text{data}}$ evaluated at $\gamma_c=\gamma_c^{\text{int}}$ and $\gamma_i=\gamma_i^{\text{int}}$.
\end{enumerate}

\begin{figure*}
    \centering
    \includegraphics[width=0.48\textwidth, height=2.8in]{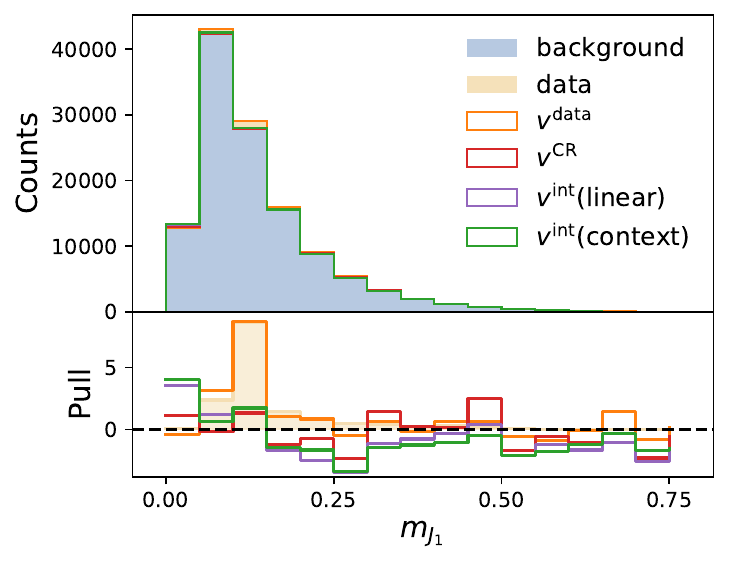}
    \includegraphics[width=0.48\textwidth,height=2.8in]{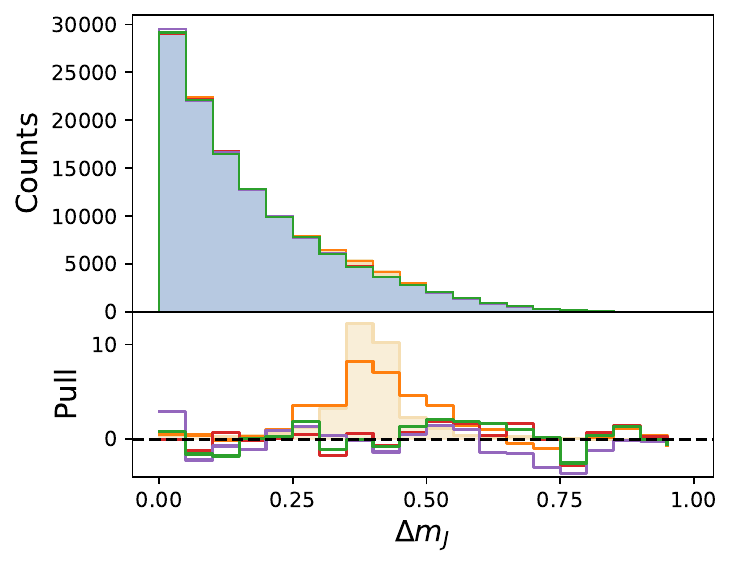}
\end{figure*}
\begin{figure*}
    \centering
    \includegraphics[width=0.48\textwidth,height=2.8in]{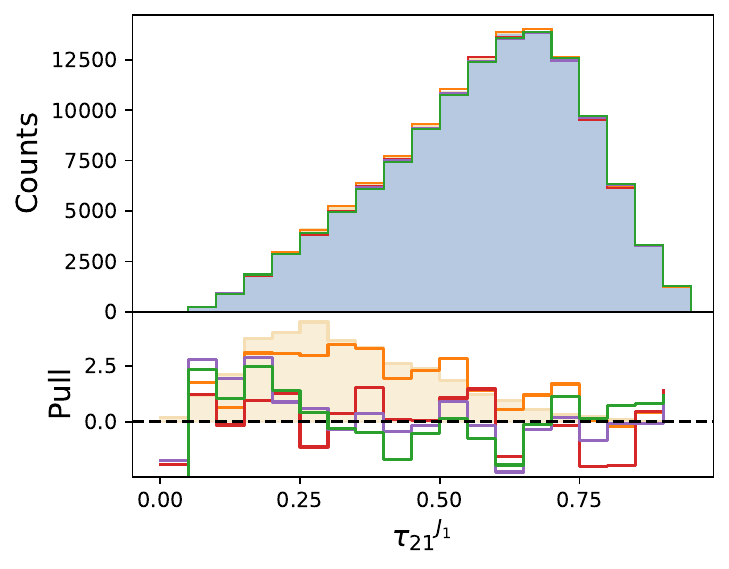}
    \includegraphics[width=0.48\textwidth,height=2.8in]{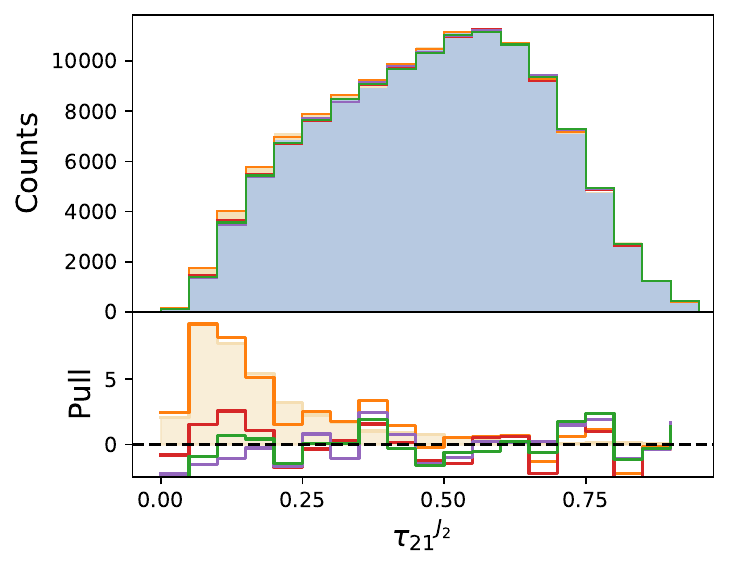}
\end{figure*}
\begin{figure*}
    \centering
    \includegraphics[width=0.48\textwidth,height=2.4in]{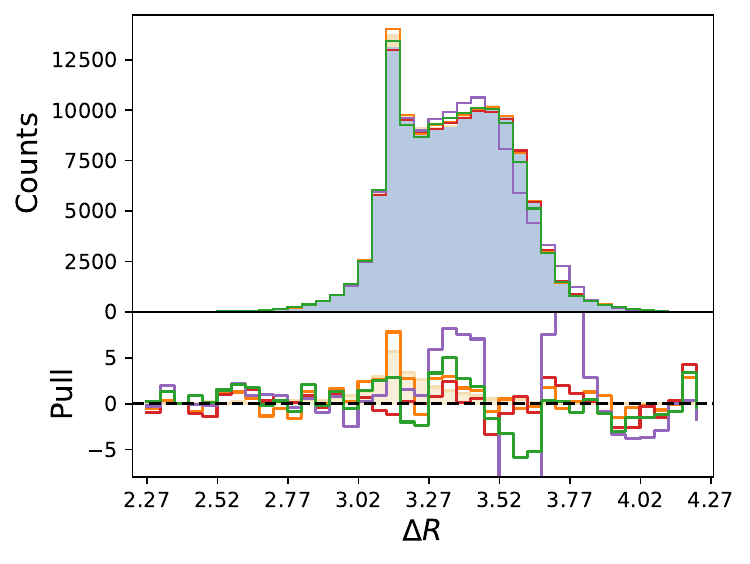}
    \caption{For $N_{sig}=3000$ and $m \in \rm SR$, we show: (Top Panel) The histograms of the samples of the features. The shaded blue histogram is the background present in the data. The shaded yellow is the signal added to the background. (Bottom Panel) The pulls with respect to the background distribution. 
    }
    \label{fig:samples}
    
\end{figure*}

\mycomment{
Using Equation~\eqref{eq:resnet_conditioning} and \eqref{eq:input_embed}, the output after 4 \texttt{ResNet} blocks is 
\begin{equation}
\begin{split}    
    v(x|t,m) = &  h^3 + \phi^4(h^3)*\gamma_c(m,t) \\
    = &  \gamma_i(x,m,t) + \left[\phi^4(h^3) + \right.\\
    & \left. \phi^3(h^2) + \phi^2(h^1)  + \phi^1(\gamma_i)\right] * \gamma_c(m,t) \\
\end{split}
\end{equation}

By doing interpolation using \eqref{eq:int_cond_embed} and \eqref{eq:int_input_embed}, we will have linearly interpolated $f^1, f^2, f^3, f^4$ and $\gamma_i$, and thereby $h^1, h^2, h^3$ and $\gamma_i$. Hence our way of interpolation is equivalent to interpolating the output of every \texttt{ResNet} Block (including the input embedding). Intuitively this means we are interpolating the $m$-dependent small changes that the \texttt{ResNet} learns, to regress $v(x|t,m)$.
}

\subsection{Model Hyperparameters and training}
\label{sec:model_hyperparams}
For training $v^{\text{data}}_{\theta}(x|t,m)$, we use $L=9$, $L'=3$. The MLPs $\phi^c$ and $\phi^i$ in \eqref{eq:cond_embed} and \eqref{eq:input_embed} consist of a single layer of $\text{dim} = 256$, with \texttt{ReLU} activation. We use 4 \texttt{ResNet} blocks, with MLPs $\phi^k$ consisting of 2 layers of dim$=256$, with $\text{dropout}=0.2$ and \texttt{BatchNormalization} in between the layers.  We train the models using \texttt{Pytorch}~\cite{pytorch} to a maximum of $2000$ epochs, with a batch-size of $4096$, using the \texttt{AdamW} optimizer \cite{adamw} with a learning rate of $3 \times 10^{-4}$, and \texttt{ReduceLROnPlateau} learning rate scheduler. After the first 100 epochs, we evaluate the negative log-likelihood (using Equation \eqref{eq:flow_density}) for the validation data every 5 epochs. The negative log-likelihood is used for selecting the best 10 models. It is also used as a metric for the learning rate scheduler, which reduces the learning rate by a factor of $0.3$ if the metric does not decrease for 50 epochs. We use early-stopping to stop the training if the validation negative log-likelihood does not improve for 100 epochs. We use the same set of hyperparameters for the CR model as well. 

To generate samples from the vector field, we first sample noise from the base distribution, which is used as the initial condition to integrate the vector field from $t=1$ to $t=0$, using \texttt{odeint} from \texttt{Zuko}~\cite{odeint}. This function uses the adaptive Dormand-Prince method~\cite{cnf, zhuang2020adaptivecheckpointadjointmethod} to perform the integration. For $\vint$, we choose $m_1=3.2$ TeV and $m_2=3.8$ TeV.

Finally, for the classifier, we use boosted decision trees (specifically a \texttt{HistGradientBoostingClassifier} from \texttt{scikit-learn}), which has been shown in \cite{cathodebdt} to be robust under noisy features. We train this BDT with default hyperparameters for 200 epochs.

\section{Results}
\label{sec:results}

\subsection{Samples}
\label{sec:samples}

In Figure~\ref{fig:samples},  we compare the samples from different models to background and data using histograms\footnote{Samples from $\vdata$ have the same normalization as the data distribution. Whereas the samples from $\vint$ (linear), $\vint$ (context) and $\vCR$ have the same normalization as the background distribution.} and pull distributions (with respect to the background)\footnote{We assume Poisson uncertainty for the pulls.}, for $N_{\text{sig}}=3000$. We find that pulls for the samples from $\vdata$ show the presence of signal, similar to the pulls for the data distribution. For the samples from different interpolation methods, higher pull values show significant deviations from the background distribution. All the interpolation methods can smoothen out the signal, but in general $\vCR$ seems to produce the best background template, followed by $\vint$ (context), and finally  $\vint$ (linear) being the worst in quality. This is especially clear in the $\Delta R$ distribution, a feature strongly correlated with $m$. Here we see the best pulls from $\vCR$, significantly exaggerated pulls from $\vint$ (linear), and moderate pull values from $\vint$ (context). 
This shows that $\vint$ (context) better models the correlation in $m$ as compared to $\vint$ (linear). In Appendix~\ref{appendix:classifier_test}, we further analyze the samples using the classifier test. In Appendix~\ref{appendix:sample_quality_SR}, we show the results for the sample quality for different SRs.

\mycomment{ su However the samples from $v^{\text{data}}_{\theta}$ for $m \in$ SR, does learn the signal to some extent, whereas after the interpolation, this signal is no longer visible in the samples from $v^{\text{interpolated}}_{\theta}$.
}

\subsection{Anomaly detection performance}

The main performance metric we use for anomaly detection is the Significance Improvement Characteristic (SIC). In terms of the signal efficiency ($\epsilon_S$ or TPR) and background efficiency ($\epsilon_B$ or FPR), one has 
\begin{equation}
\text{SIC} = {\epsilon_S\over\sqrt{\epsilon_B}}.
\end{equation}
The SIC characterizes the improvement to the nominal significance, using a cut on the anomaly score from a given method. SIC is dependent on the working point, and we present this dependence by plotting SIC vs.\ the background rejection ($1/\epsilon_B$).
At lower signal efficiencies, there are large statistical uncertainties in the SIC values, so we introduce a cut-off in the SIC curves when the relative statistical error on the background efficiency exceeds 20\%. We also compare SIC at  $\text{FPR}=0.001$ to represent a typical, fixed working point that an experiment might choose in practice.

\begin{figure*}
    \centering
    \includegraphics[width=0.48\textwidth]{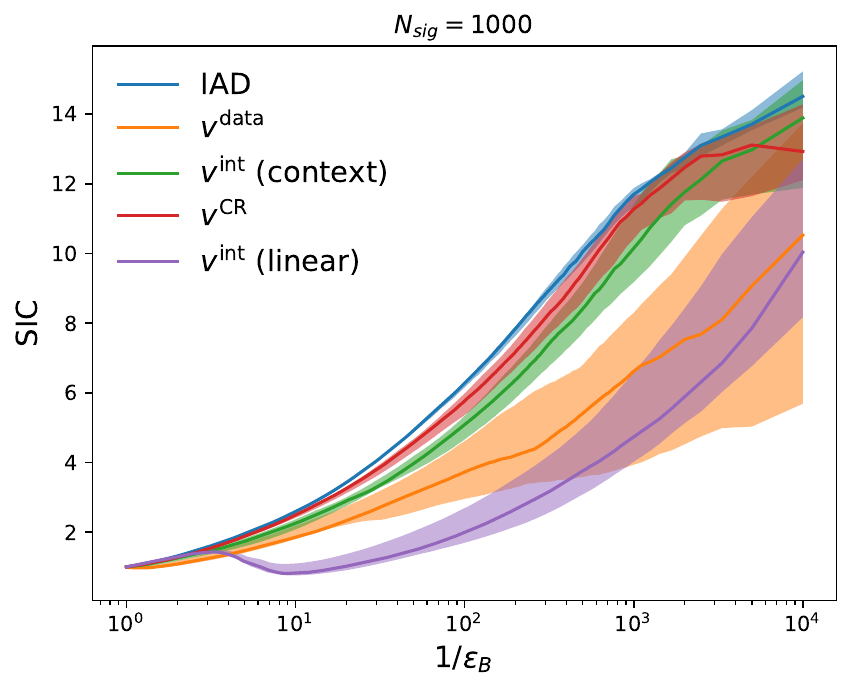}
    \includegraphics[width=0.48\textwidth]{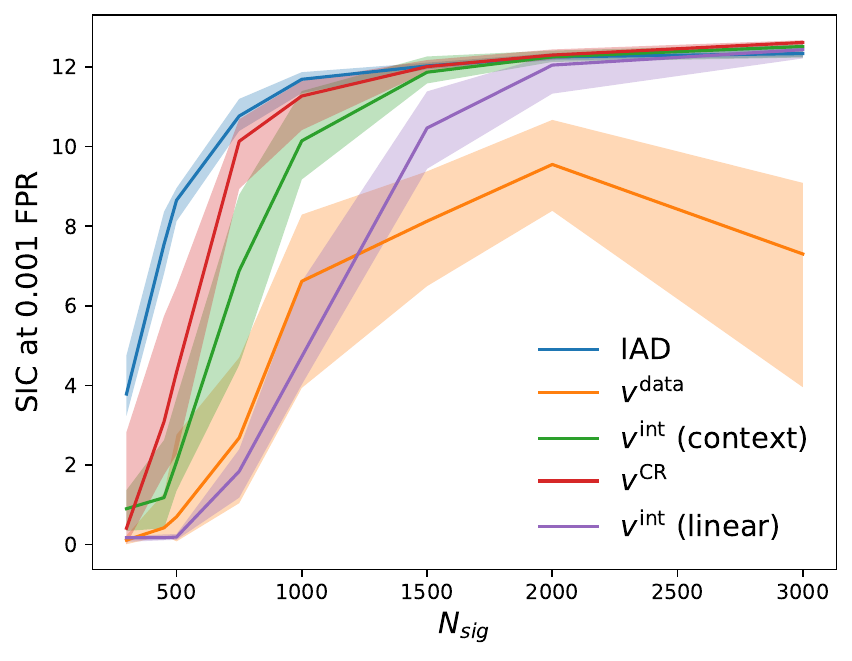}
    \caption{(Left) SIC curves for a signal injection of $1000$ (which corresponds to $S/\sqrt{B} \approx 2.2$). (Right) SIC at $\epsilon_B=0.001$ for different signal injections. While  $\vCR$ retains the best signal sensitivity, the much more computationally efficient $\vint$ (context) is only a little bit worse by comparison. Meanwhile, $\vint$ (linear) shows considerably worse performance at lower signal strengths. }
    \label{fig:sic_1000}
\end{figure*}

In Fig.~\ref{fig:sic_1000} (left), we compare the SIC curves for $N_{sig}=1000$, showing error bars for 10 different signal injections. The SIC curves for samples from $v^{\text{data}}_{\theta}(x|t,m)$ have large error bars. This is a result of the model being able to learn the full data distribution better for some signal injections (resulting in worse SIC), and so not so well in other cases (resulting in better SIC)\footnote{One could possibly improve this error bar by ensembling multiple generative models on the same dataset, which we leave for future studies.}.  

The weakly supervised classifier identifies both the signal and the mismodeling artifacts, evident in the pull distributions in Figure~\ref{fig:samples}. At higher signal strengths the classifier is able to better identify the signal along with the mismodeling effects. At lower signal strengths, the mismodeling effects overcome the signal, which worsens the signal sensitivity. Hence $\vCR$, with the best pulls in Figure~\ref{fig:samples}, results in the best SIC at all signal strengths in Figure~\ref{fig:sic_1000} (Right). Both $\vint$ methods are able to interpolate out the signal, leading to an improvement over $\vdata$ for $N_{sig} > 1500$. At lower $N_{sig}$ values, the performance of $\vint$ (context) hews fairly closely to the benchmark set by $\vCR$. This is significantly better than $\vint$ (linear), whose performance suffers much more, due to the worse pulls shown in Figure~\ref{fig:samples}.  


\subsection{Timing Comparison}
In Table~\ref{tab:method_comparison}, we compare the total time taken for generating background templates for different methods. The background template generation method in \cite{cathode, anode, lacathode, ranode} uses normalizing flows, which takes 3 hours to train. The flow-matching version introduced in this paper, $\vCR$, offers a significant speedup to around to 30 mins per SR. The training time for the base model of CURTAINS4F4, which also uses normalizing flows, is around 3 hours, and the lighter flow for each SR trains in 7 mins~\cite{curtrainsf4f}. Using optimal transport, RAD-OT offered a faster alternative to CURTAINS4F4 at 10 mins per SR. Finally, our method SIGMA, trains the initial flow-matching model for 30 mins and subsequent sampling takes 30 seconds per SR. Hence SIGMA offers significant gains in computational efficiency compared to other methods.

\begin{table}[h!]
\centering
\renewcommand{\arraystretch}{1.4} 
\begin{tabular}{p{3.2cm} p{2.9cm} p{2.8cm}}
\hline
\textbf{Method}          & \textbf{Model} & \textbf{Timing} \\ 
\hline
CATHODE/ANODE            & Normalizing Flows         & 3 hours per SR  \\ 
CATHODE/ANODE            & Flow Matching             & 30 mins per SR  \\ 
CURTAINS4F4              & Normalizing Flows         & 3 hours \\
                         &                           & (base model)   \\ 
                         &                           & +7 mins per SR   \\ 
RAD-OT                   & Optimal Transport         & 10 mins per SR  \\ 
\textbf{SIGMA}           & \textbf{Flow Matching}    & \textbf{30 mins} \\
                         &                           & \textbf{(training)}   \\ 
                         &                           & \textbf{+30 secs per SR}  \\
\hline
\end{tabular}
\caption{Timing comparison of different methods.}
\label{tab:method_comparison}
\end{table}

\mycomment{The interpolated results have better performance along with tighter error-bars, which shows that $v^{\text{int}}_{\theta}(x|t,m)$ is able smooth out whatever $v^{\text{data}}_{\theta}(x|t,m)$ has learned. This is also apparent in Fig~\ref{fig:sic_1000}. We also see in Fig~\ref{fig:sic_1000} that this new interpolation has almost similar performance to $v^{\text{CR}}_{\theta}(x|t,m)$. Hence we see that this new way of interpolation retains the anomaly detection performance, while reducing the computational cost.}

\mycomment{
One can think of numerous possible ways of doing this interpolation, instead of Equation~\eqref{eq:int_cond_embed},~\eqref{eq:int_input_embed}. We observed that if only parts of the model is interpolated in $m$, whereas the other parts depend on $m \in \rm SR$, the performance is worse. This is because the model possibly still retains information about the signal through its dependence on $m \in \rm SR$. One could also do an interpolation after a full forward pass, which is equivalent to doing a linear interpolation in the vector field:
\begin{equation}
\label{eq:linear_interp_vector}
\begin{split}
    v^{\text{linear interp}}_{\theta}(x|t,m) & = w \times \vdata (x|t,m_1) \\ 
    & + (1-w) \times \vdata (x|t,m_2).
\end{split}
\end{equation}
One could show using \eqref{eq:flow_density}, that this is equivalent to linearly interpolating the log density in $m$. We found that this results in much worse background templates, with or without a frequency embedding (see Figure.~\ref{fig:linear_interp_vector} for the result with frequency embedding).
}

\section{Conclusions}
\label{sec:conclusions}

In this work we developed SIGMA, a new and efficient interpolation method for generative modeling of backgrounds in resonant anomaly detection. Rather than needing to train separate generative models in the complement of each signal region, SIGMA re-uses a single generative model trained on all of the data, with a subsequent interpolation of its parameters from control regions into the signal region of interest. This reduces the computational cost of SIGMA significantly relative to previous approaches such as ANODE and CATHODE \cite{anode, cathode}, while preserving the high quality of their background templates and signal sensitivity.

The question of choosing the best interpolated background template in a data-driven way is an open one. We showed that the SIC, described in Section~\ref{sec:results}, is very sensitive to bad background templates. For the LHCO R\&D dataset, the $\vint$ (context) method showed the best performance, only a little worse than $\vCR$. However, with real data we obviously do not have access to signal and background labels to calculate the SIC with. Nevertheless, it should be possible to perform signal injection tests, as ATLAS and CMS currently do to set limits on signal models \cite{ATLAS:2020iwa,CMS-PAS-EXO-22-026}, with artificial gaussian or other signals added to real data. One could then see first-hand whether the ``all data" model learns the data plus artificial signal, and whether the interpolated versions correctly smooth out the artificial signal.



There are many interesting future directions suggested by our work. 
Given that there was still a small performance gap between the previous, expensive interpolation method (masking out the CR) and SIGMA, it is worthwhile to explore further improvements to our method, such as using diffusion models instead of flow-matching or performing some kind of non-linear interpolation using more than two mass points in the control region.

It would also be interesting to see how SIGMA applies to noisy features introduced in \cite{cathodebdt}. Finally, it should be totally straightforward to apply SIGMA to higher-dimensional SRs (multiple resonant features), which previously would have been even more computationally prohibitive. 



\section*{Acknowledgments}

We are grateful to  Darius Faroughy and Sung Hak Lim for helpful discussions and Gregor Kasieczka for comments on the draft. This work was supported by DOE grant DE-SC0010008. This research used resources of the National Energy Research
Scientific Computing Center, a DOE Office of Science User Facility
supported by the Office of Science of the U.S. Department of Energy
under Contract No. DE-AC02-05CH11231 using NERSC award
HEP-ERCAP0027491. The authors acknowledge the Office of Advanced Research Computing (OARC) at Rutgers, The State University of New Jersey \url{https://it.rutgers.edu/oarc} for providing access to the Amarel cluster and associated research computing resources that have contributed to the results reported here.

\appendix

\section{Importance of using frequency embedding}
\label{appendix:freq_embedding}

The frequency embedding is necessary for $\vdata$ to learn the small signal component in data. In this discussion, we denote the data model without a frequency embedding (i.e., using $\alpha(m)=m$ with $m$ being an unscaled mass) as $\vtildata$. The data model with a frequency embedding, given by Equation~\eqref{eq:mass_embed}, is denoted as $\vdata$.

\begin{figure*}
    \centering
    \includegraphics[width=0.48\linewidth]{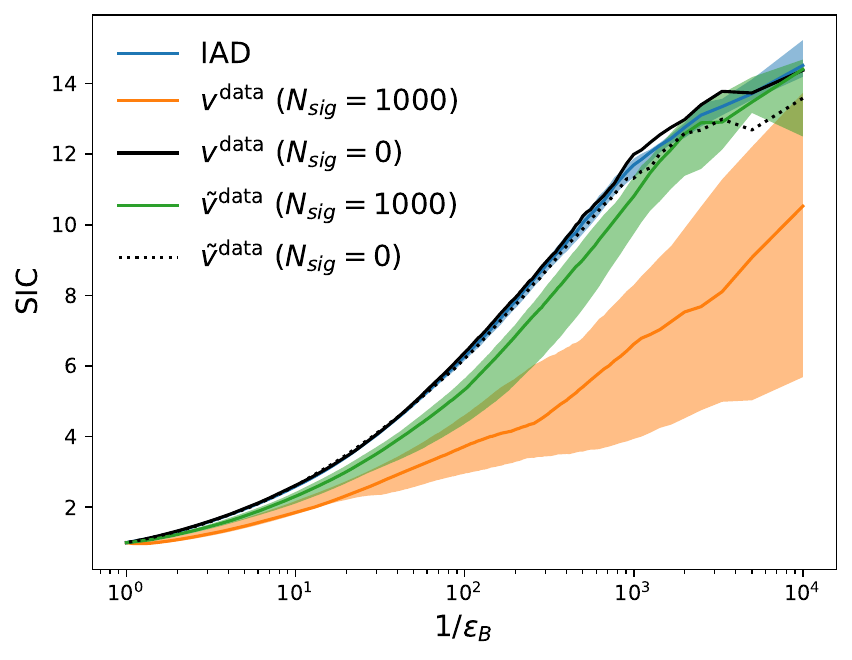} 
    \caption{The SIC curve for background template obtained $\vdata$ (in black, solid) and $\vtildata$ (in black, dotted), trained on data without signal injection are similar to IAD.  However, comparing the same models when trained with signal injection, show that the SIC for $\vdata$ (with frequency embedding), is worse than $\vtildata$ (without frequency embedding). This is because $\vdata$ learns the small signal component in data, better than $\vtildata$. Hence, the frequency embedding is crucial for the model to best learn the data.}
    \label{fig:embedding}

\end{figure*}

As seen in Figure~\ref{fig:embedding}, without signal injection, both $\vdata$ and $\vtildata$ give a SIC very close to the IAD. With the signal injection, this SIC worsens for both models. $\vdata$ learns the signal better and hence shows worse SIC values than $\vtildata$.

For our method to be fully signal-model agnostic, we do not want the data model to smoothen out parts of the data distribution that could be relevant to the background template. Hence it is crucial that we start with the best possible data model, which in this case is the model with a frequency embedding.

\section{Classifier Test}
\label{appendix:classifier_test}
To better understand what the models learned, we study the weight distributions from the classifier test introduced in \cite{das2023understandlimitationsgenerativenetworks}. We train an ensemble of 50 \texttt{HistGradientBoostingClassifier} classifiers to distinguish the samples from each model vs the true background samples. Hence the weight distributions show: 
\begin{equation}
    \text{weight}(x) = \frac{p_{\text{samples}}(x)}{p_{\text{background}}(x)}.   
\end{equation}
If model $v^{\text{data}}_{\theta}$ does learn the signal, we expect the samples after a cut $\text{weight}(x)>1$ to look like the signal. Whereas a similar cut on samples from $v^{\text{CR}}_{\theta}$ or  $v^{\text{int}}_{\theta}$ should not look like the signal and should correspond to modeling between the interpolated samples and the background. 

\begin{figure*}[h]
    \centering
    \includegraphics[width=0.48\textwidth]{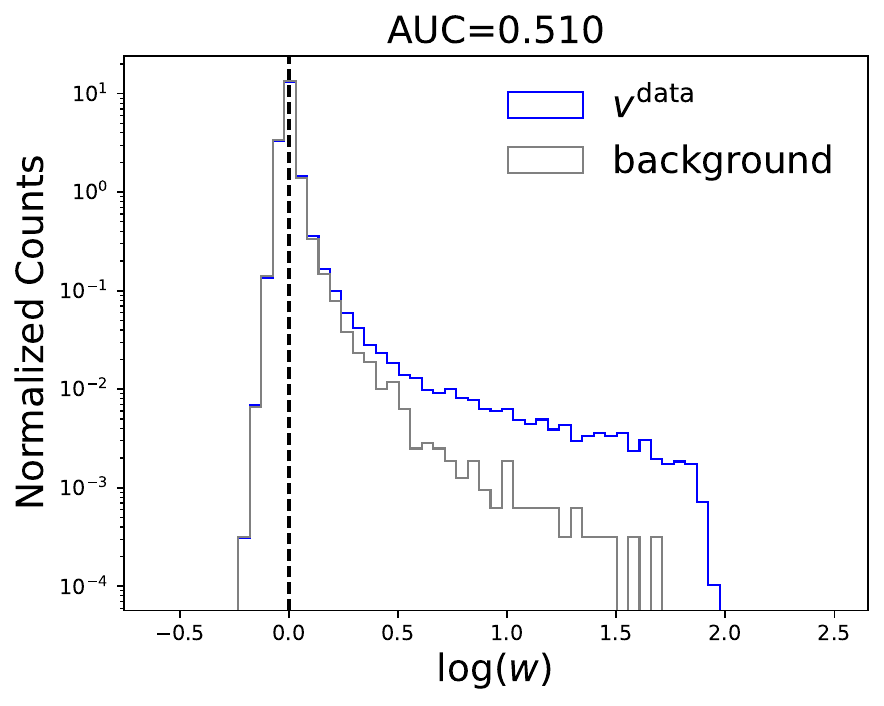}
    \includegraphics[width=0.48\textwidth]{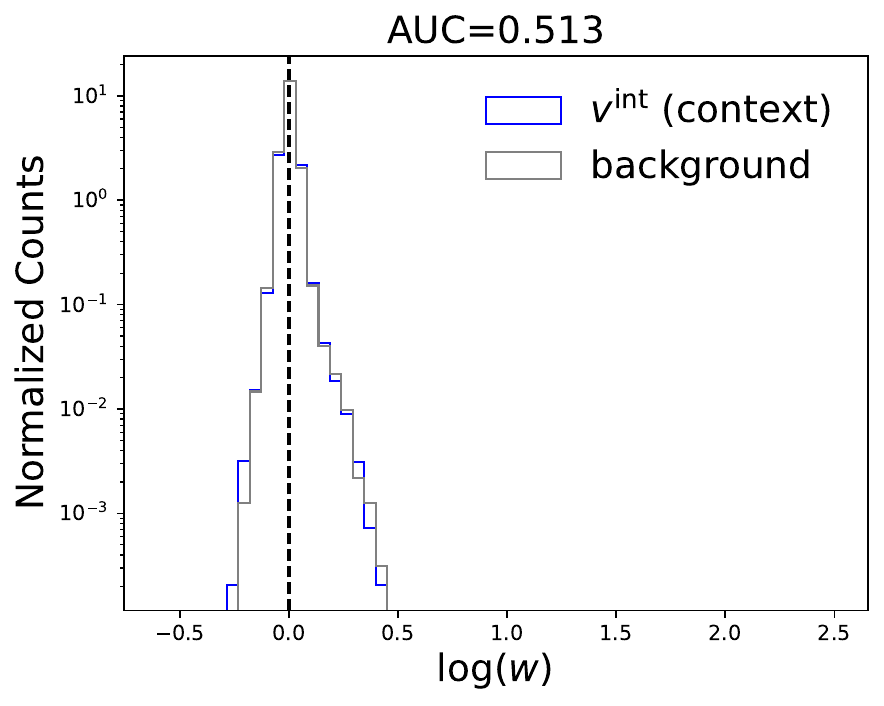}
    \includegraphics[width=0.48\textwidth]{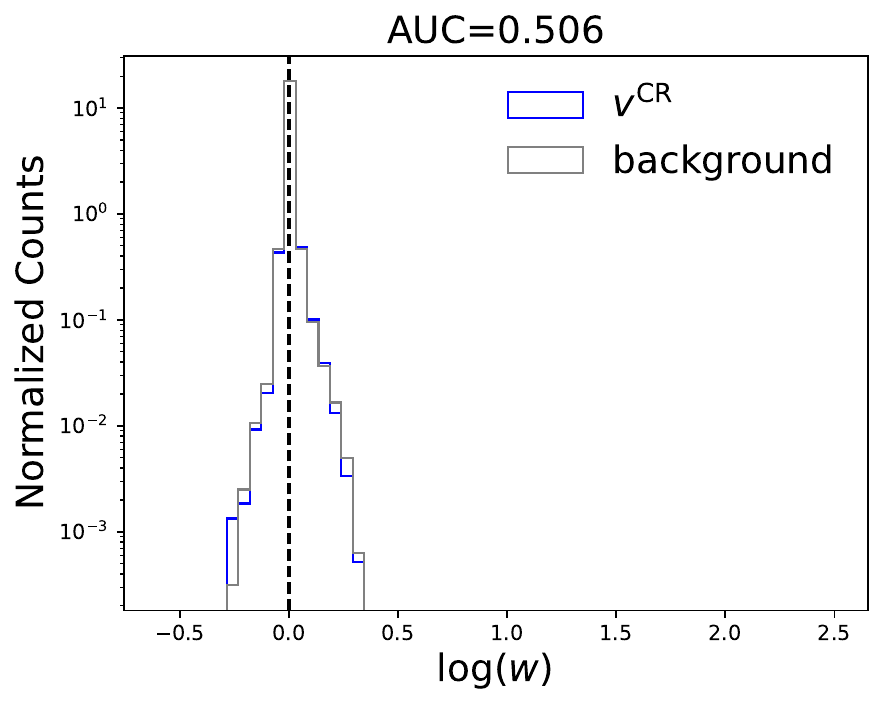}
    \includegraphics[width=0.48\textwidth]{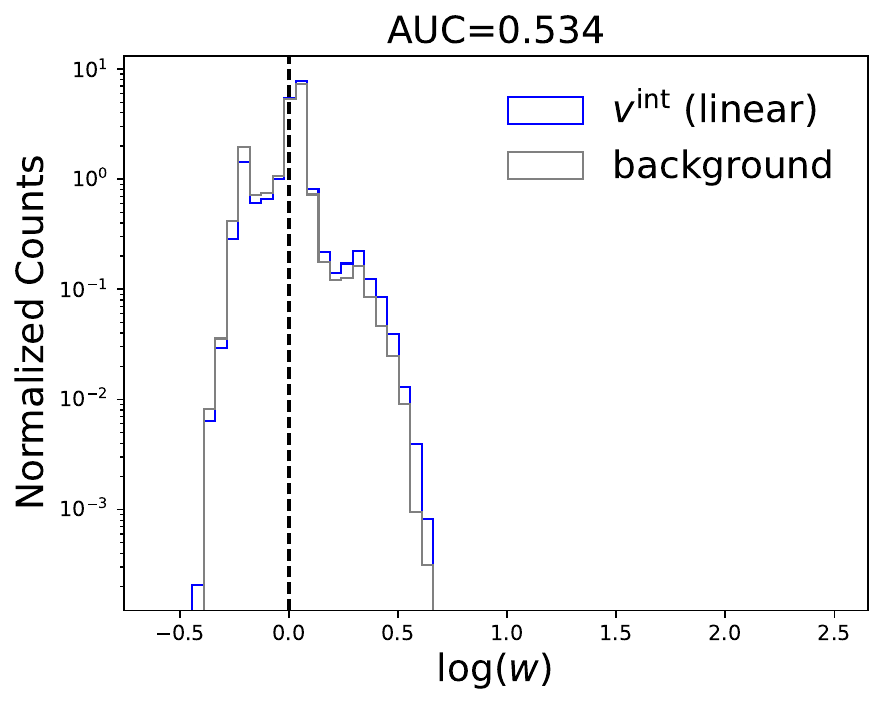}
    \caption{The weight distributions for classifier trained on samples from $v^{\text{data}}$ (Top Left)/ $v^{\text{CR}}$ (Bottom Left)/ $v^{\text{int}}$ (context) (Top Right)/ $v^{\text{int}}$ (linear) (Bottom Right) vs true background. The AUC of $\vint$ (context) is slightly worse than $\vdata$, which produces a long tail for $\log (\text{weight}) > 0$ corresponding to the signal.} 
    \label{fig:weight_dist}. 

\end{figure*}

\begin{figure*}
    \centering
    \includegraphics[width=0.48\textwidth]{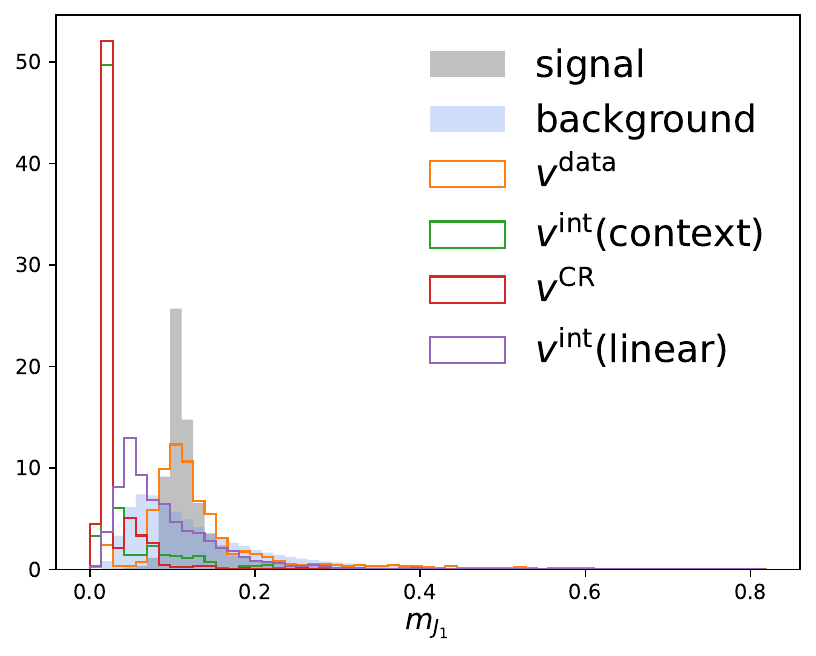}
    \includegraphics[width=0.48\textwidth]{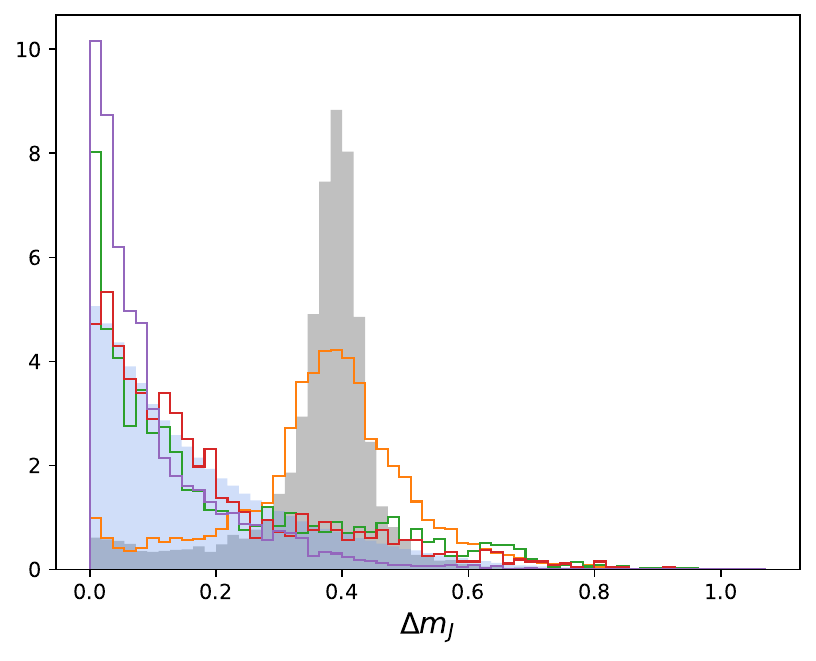}
    \includegraphics[width=0.48\textwidth]{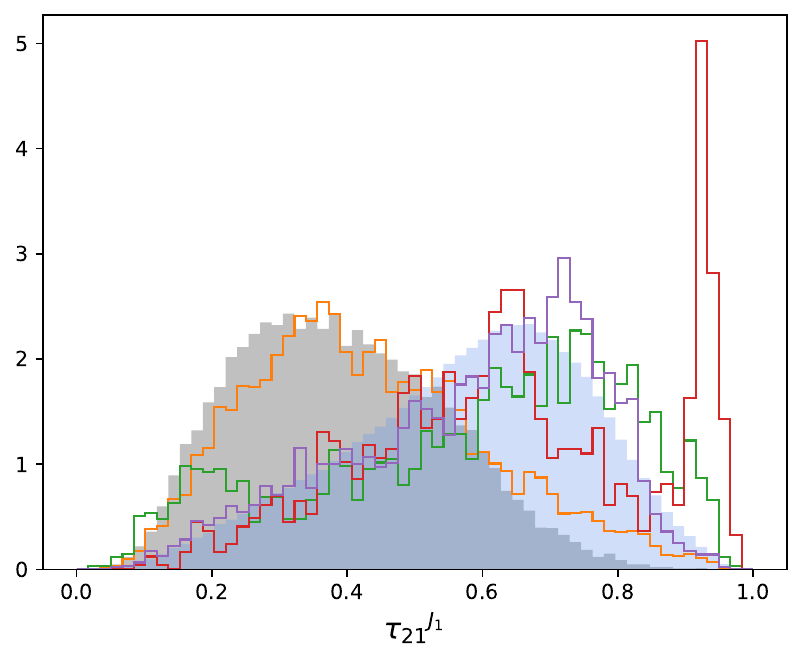}
    \includegraphics[width=0.48\textwidth]{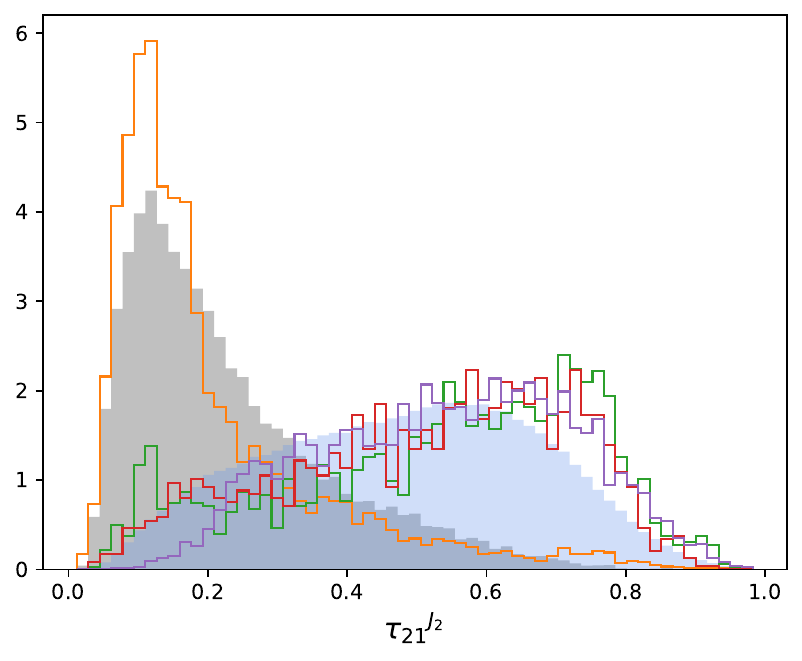}
    \includegraphics[width=0.48\textwidth]{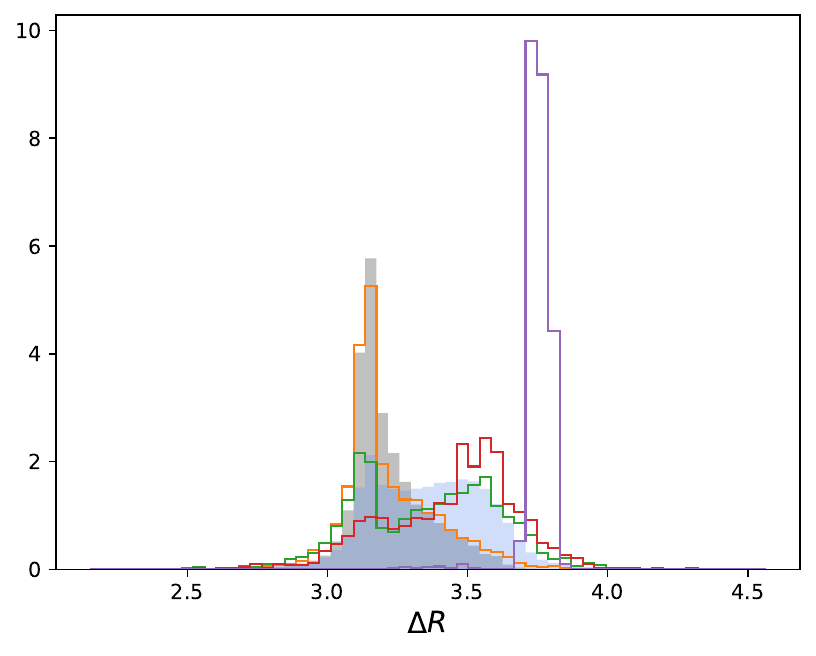}
    \caption{For $N_{sig}=3000$ and $m \in \text{SR}$, this figure shows the histograms of the samples of the features after a classifier cut $\log \text{weight} > 0.1$. For $\Delta R$, we chose a higher cut $\log \text{weight} > 0.2$. After a cut, we see that the samples from $v^{\text{data}}$ (in orange) are localized around the signal. This shows that the model was able to learn the signal. The samples from $\vint$ (context), $\vint$ (linear) and  $\vCR$ do not show the signal. The samples from $\vint$ after a cut look similar to the other background templates, except for $\Delta R$, where the classifier finds the mismodeling artifacts. }
    \label{fig:cut_samples}
\end{figure*}

We see that $\vCR$ has the best AUC, and $\vint$ (linear) has the worst AUC. $\vint$ (context) and $\vdata$ has similar AUCs, but in Fig.~\ref{fig:weight_dist} for samples from $v^{\text{data}}_{\theta}$, the weight distribution has a long tail for $\text{weight}>1$. The resulting samples from $v^{\text{data}}_{\theta}$ after a cut $\log (\text{weight})>0.1$ in Figure~\ref{fig:cut_samples} show the peaks where the signal is localized, as expected. The interpolated samples after the same cut don't show the signal.\footnote{Samples with similar AUCs result in samples with different anomaly detection performance. This shows that AUC by itself can be misleading, and one needs to study the weight distributions to get a better understanding, as was shown in  \cite{das2023understandlimitationsgenerativenetworks}.} This shows that $v^{\text{int}}_{\theta}$ is able to remove the signal. For $\Delta R$ samples after a cut, $\vint$ (linear) shows the mismodeling artifacts which are also visible in pull distributions in Figure~\ref{fig:samples}.

\begin{figure}[H]
    \centering
    \includegraphics[width=0.9\linewidth]{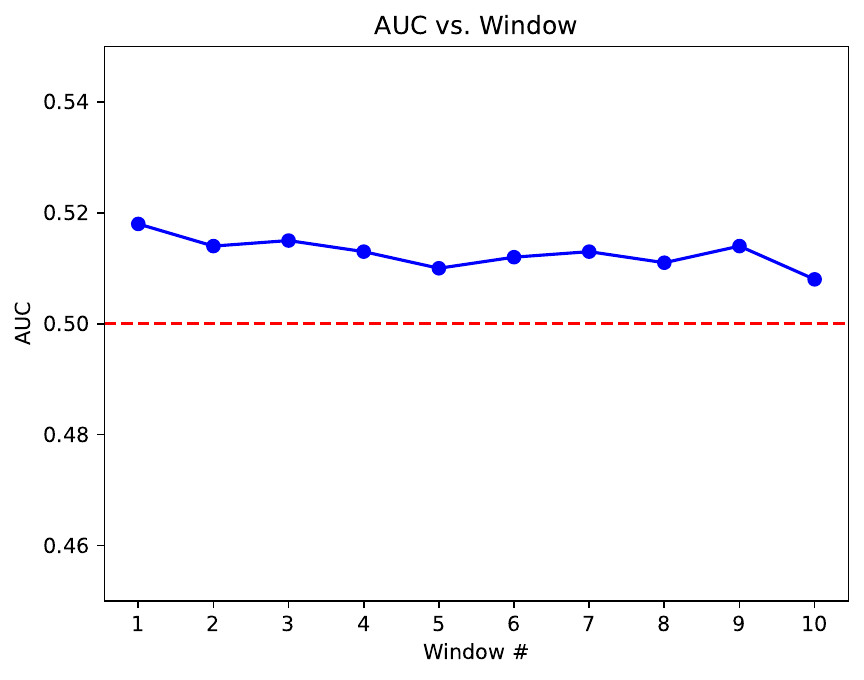}
    \caption{The figure shows the classifier AUC for background template quality in Signal Regions centered at $m_{JJ} = \left(3.5 - 0.1 \cdot (5 - n)\right)~\text{TeV}$, for $n = 1, \ldots, 10$. Classifiers are trained between true background samples in SR, and background template obtained from $\vint$ (context). SIGMA is robust across different SRs in the resonant feature.}
    \label{fig:otherSR}
\end{figure}

\section{Sample quality across different SRs}
\label{appendix:sample_quality_SR}

In the main body of the paper we focused on the SR $m_{JJ}\in [3.3,3.7]$~TeV that is centered on the LHCO R\&D signal with $m_{JJ} = 3.5$~TeV. Here we explore the  quality of generated samples interpolating into other SRs that do not contain the signal. Taking the same FM model trained on the data which contains 1k signal events as in the main body of the paper, we interpolated it into overlapping SRs of width $400$ GeV, centered at $m_{JJ} = \left(3.5 - 0.1 \cdot (5 - n)\right)~\text{TeV}$, for $n = 1, \ldots, 10$. $n=5$ corresponds to the SR considered in the main body of the paper.

For each SR, we trained classifiers between the true background samples (in SR) vs the background template obtained from $\vint$ (context). We used the classifier AUC as a metric to assess the background template quality.
As shown in Figure~\ref{fig:otherSR}, we find that the classifier AUCs for all the SRs are close to $ \sim 0.5$. This shows that SIGMA is robust across different SRs.

\bibliographystyle{apsrev4-1}

\bibliography{references}

\end{document}